\documentclass[12pt]{iopart}

\pdfoutput=1 
\usepackage{ifpdf}
\usepackage{iopams}
\usepackage{graphicx}

\begin{document}

\title{Perturbations of the local gravity field due to mass distribution on precise measuring instruments: a numerical method applied to a cold atom gravimeter}

\author{G. D'Agostino}
\address{INRIM, Istituto Nazionale di Ricerca Metrologica, 73 Strada delle Cacce, 10135 Turin, Italy}

\author{S.~Merlet, A.~Landragin and F.~Pereira Dos Santos}
\address{LNE-SYRTE, Observatoire de Paris, CNRS et UPMC, 61 avenue de l'Observatoire, 75014 Paris, France}
\ead{sebastien.merlet@obspm.fr}

\begin{abstract}

We present a numerical method, based on a FEM simulation, for the determination of the gravitational field generated by massive objects, whatever geometry and space mass density they have. The method was applied for the determination of the self gravity effect of an absolute cold atom gravimeter which aims at a relative uncertainty of $10^{-9}$. The deduced bias, calculated with a perturbative treatment, is finally presented. The perturbation reaches $(1.3\pm0.1)\times10^{-9}$ of the Earth's gravitational field.

\end{abstract}

\maketitle

\section{Introduction}\label{intro}

Gravitational forces from surrounding masses are usually negligible compared to electro-magnetic forces and thus not considered as relevant in laboratory based experiments. Nevertheless, they can significantly impact on accurate forces measurement~\cite{Kibble1976} or even constitute the dominant effect~\cite{Gundlach2000, Fixler2007}. As an example, an aluminum disk with $50~$cm diameter and $4~$cm thickness generates a gravitational field ranging from 4 to 1 parts in $10^9$ of the Earth's gravity field $g$, from its surface to $30~$cm along its axis. Components with similar mass density and dimensions are commonly used in precise instruments developed for measuring the gravity field with uncertainties of few parts in $10^9$. The measurement principle of these instruments is based on the determination of the ballistic trajectory followed by a mass, i.e. corner cube retro-reflectors and recently atoms, during their vertical free-fall under the influence of gravity \cite{Niebauer1995, Agostino2008, Merlet2010}. Due to constrains in their design, their mass distribution is in general not symmetric with respect to the trajectory of the test mass. The effect of parasitic attractions which is called self gravity effect when restricted to the influence of the device itself, has to be carefully estimated and if necessary corrected for in order to guarantee the accuracy of the measurement.

In the case of the absolute corner cube gravimeter FG5, authors in~\cite{Niebauer1995} indicate that the $g$ measurement is corrected from the attraction of the apparatus, with an uncertainty of $10^{-9}~$m.s$^{-2}$ without giving details on its calculation. A similar level of uncertainty is also necessary in the case of the cold atom gravimeter (CAG) described in$~$\cite{Louchet-Chauvet2011}, currently operating and improved within the framework of the LNE watt balance experiment$~$\cite{Geneves2005, Merlet2008}.

In this paper we describe a numerical method for computing the gravitational field generated by extended and continuous massive objects, whatever geometry and space mass density they have. Application of the method for the quantitative estimate of the self gravity error on the CAG is also given.

\section{Analogy between gravitational and electrical interactions}\label{section1}

Mass and electric charge are respectively the sources of the gravitation and electromagnetic interaction between bodies$~$\cite{Bartlett2004}. In particular the Newton's and Coulomb's laws which describe respectively the gravitational and electric force occurring between two masses and two charges, have the same behavior. This analogy can be exploited to compute gravitational forces using methods originally developed for electro-magnetic forces, by replacing individual charges by individual masses, space charge density by  space mass density and $1/4\pi\epsilon_{0}$ by $G$ where $\epsilon_{0}$ is the vacuum permittivity and $G$ is the gravitational constant.

In general, the gravitational field can be computed by solving cumbersome integrals of three-dimensional vector functions. Analytical solutions exist only for axial-symmetric and homogeneous bodies, e.g. a spherical shell, a solid sphere, a right rectangular prism, a right polygonal prism and a polyhedron~\cite{Li1998}. For more complex systems, which can not be easily broken down in simplified parts, the Finite Element Method (FEM) helps in finding an approximate numerical solution of the integrals. The essential characteristic of this method is the mesh discretization of a continuous domain in a set of discrete sub-domains.
For this application we used the FEM software packages developed by COMSOL Multiphysics~\cite{comsol}, in particular the electro-static module.

\section{Finite element method and expected accuracy}\label{section2}

In our application, the three-dimensional model of the object in a $Oxyz$ frame of reference defines a body domain $BD$ filled-in a given mass density, whereas the surrounding domain $SD$ shapes a vacuum space where we are interested in calculating the gravitational field. The interface surface $S_{I}$ between $BD$ and $SD$ corresponds to the body geometry while the external surface $S_{E}$ limiting $SD$ depends on the boundary conditions.

\begin{figure}[!t]
\includegraphics[width=5in,bb=-150 220 400 600]{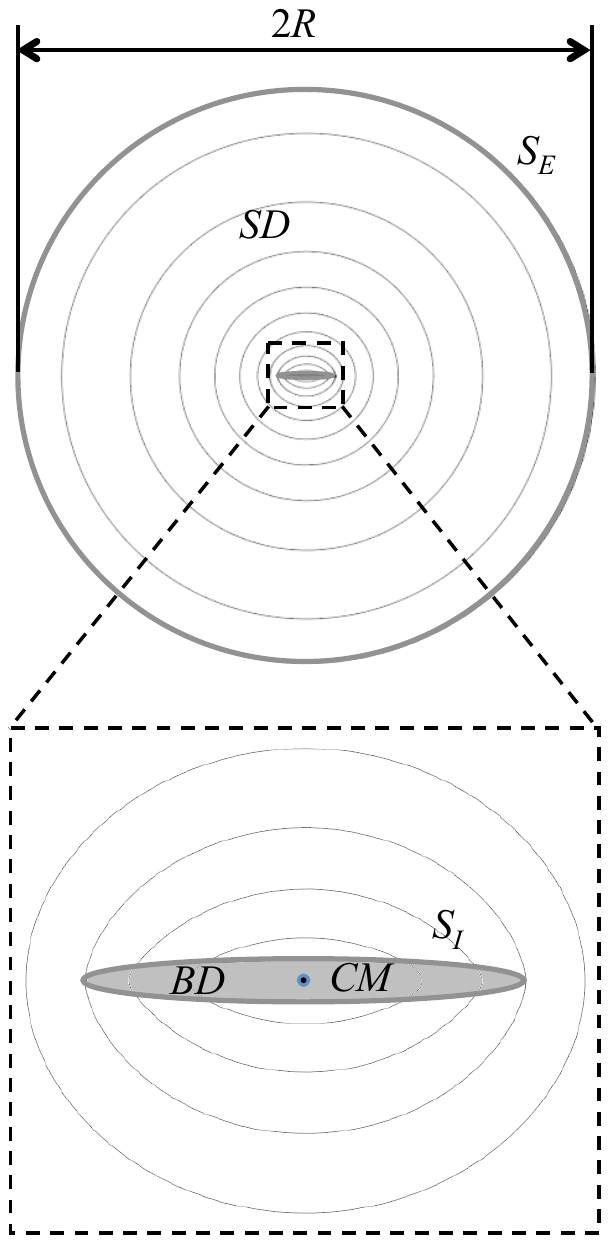}
\caption{Equipotential surfaces generated by an ellipsoidal body in its surrounding.  At sufficiently large distances from the object, these equipotential are spherical. }
\label{figfig1}
\end{figure}

Assigning the equipotential condition to $S_{E}$ simplifies the problem. Indeed, equipotential surfaces are strongly dependent on the body geometry but at infinite distance they become spheres centered on the center of mass $CM$ (see Figure~\ref{figfig1} for an example concerning an ellipsoidal body). Knowledge about the geometry of equipotential surfaces near complex objects is usually poor and the extension of $SD$ to infinity is not numerically possible. Nevertheless, defining $S_{E}$ by a sphere centered on $CM$ is the best choice, independently on the object. The systematic error arising from the possible field distortion can be minimized by increasing the sphere radius. The upper limit in stepping up the radius of $S_{E}$ is set by the finite number of mesh elements allowed by the memory of the solver.

In fact, increasing the volume of $SD$ results in a higher scattering of the solution due to a lower discretization of the space where the gravity field is computed. In conclusion the final uncertainty depends on a compromise between the bias due to the boundary condition and the scattering due to the mesh size.

COMSOL Multiphysics offers several ways to set the mesh size. In case of objects having hollow or highly sharp geometries the algorithm must resolve the gravitational filed in great detail only on some portions of the domain. In this situation best results can be achieved using the adaptive mesh generation. After a first solution obtained with the mesh size selected by the user, the algorithm provides a second solution after optimizing the mesh size in those regions requiring an higher resolution. Moreover, the ultimate accuracy compatible with the memory of the solver is achieved by constraining an extra-fine mesh in those regions where the solution is required.

To evaluate the expected uncertainty of the FEM method, a comparison was carried out between the analytical solution and the numerical results of a simulation of the same problem. LetÍs consider a right rectangular prism with uniform density and centered at the origin of a $Oxyz$ frame of reference, with the $z$ axis oriented upwards. Sides of the prism are parallel to $x$, $y$, $z$ axes with dimensions equal to $2L$, $2L$ and $L$, respectively.

\begin{figure}[!t]
\includegraphics[width=6.5in,bb=50 40 1300 600]{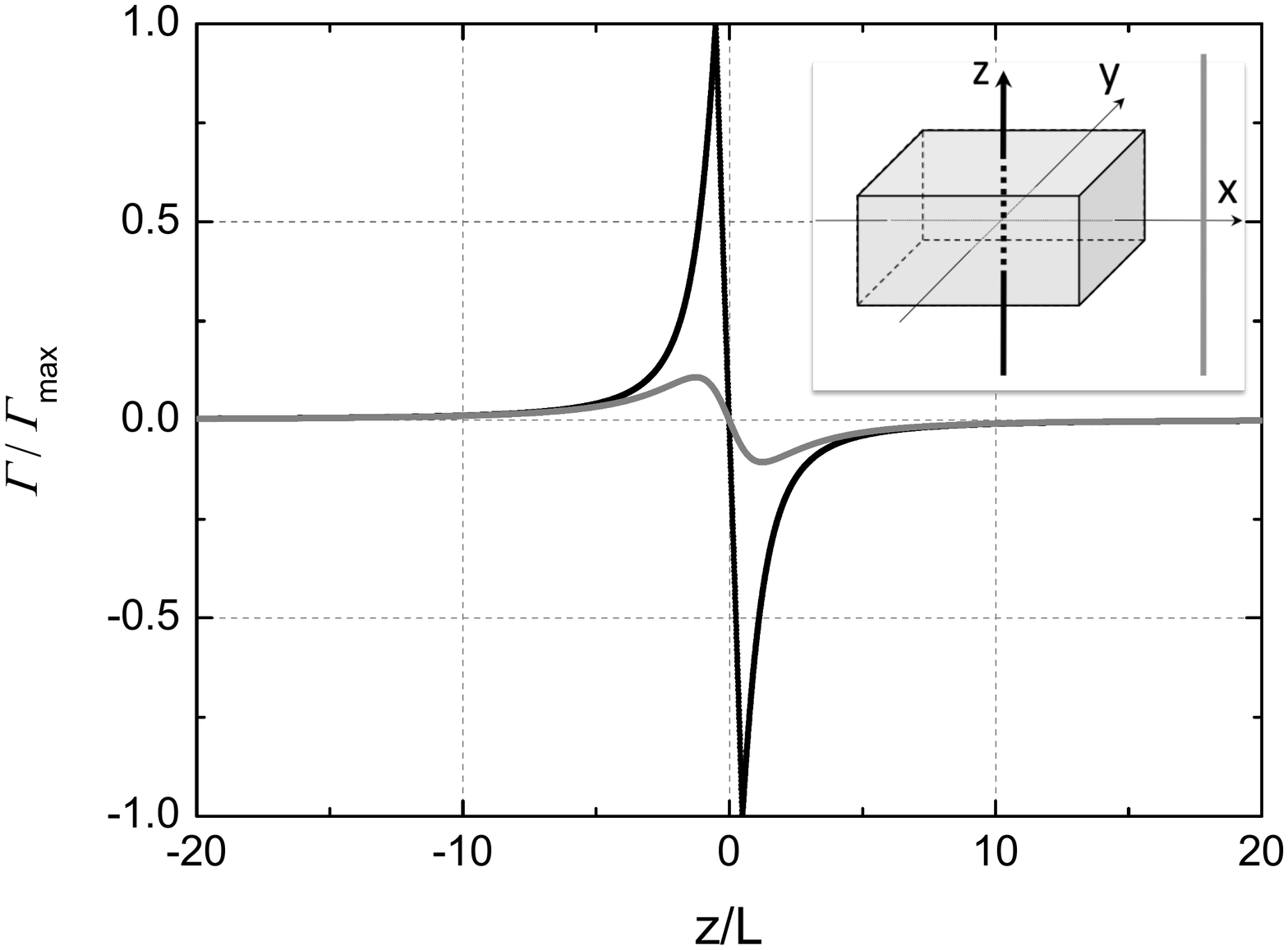}
\caption{Gravitational field of a prism of dimensions $2L, 2L, L$, calculated with the FEM simulation, along two lines parallel to the $z$ axis passing at the center of the prism (black line) and at the distance $L$ of the prism (gray line).  }
\label{figfig2a}
\end{figure}

\begin{figure}[!t]
\includegraphics[width=15in,bb=-20 20 1300 250]{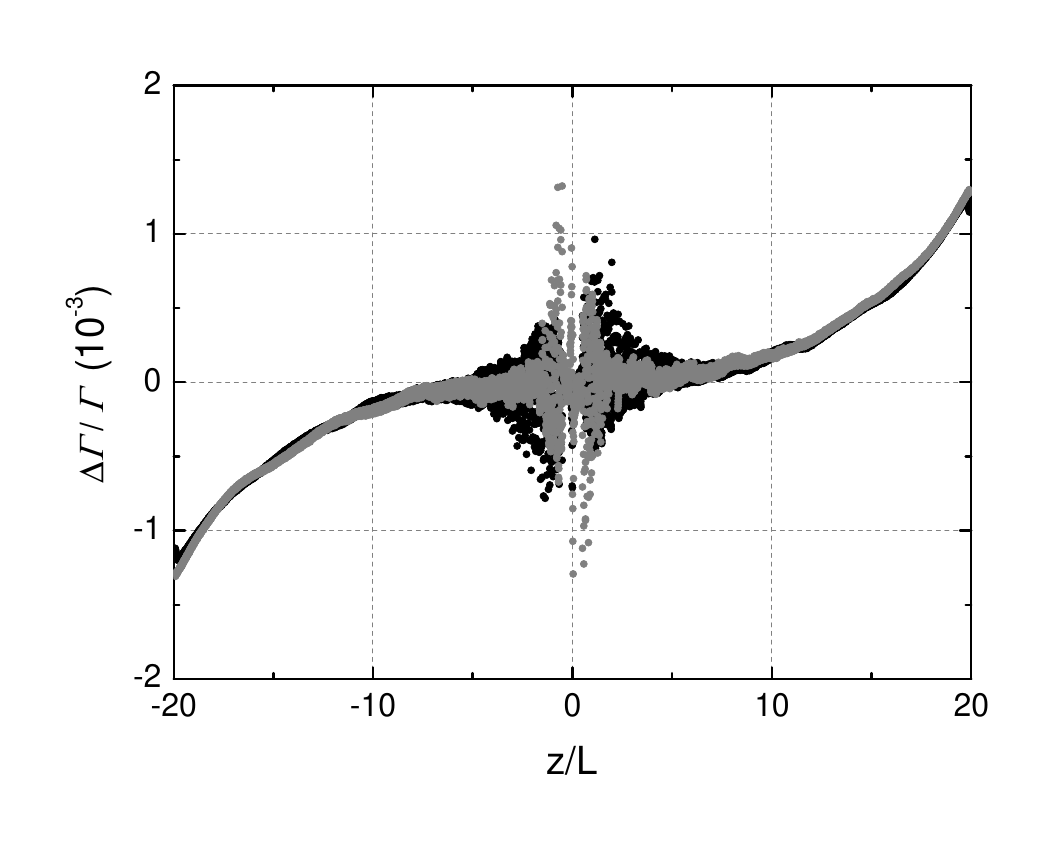}
\caption{Relative difference results between the FEM simulation of a prism ($2L, 2L, L$) represented in Figure~\ref{figfig2a}, and the analytical solution according to equation~\ref{eq3} for the two vertical lines passing at the center of the prism (in black) and  $L$ of the prism (in gray).  }
\label{figfig2b}
\end{figure}

The $z$ component of the gravitational field generated at a point ($x, y, z$) is~\cite{Li1998, Haaz1953}:

\begin{equation}\label{eq3}
\Gamma=-G\rho_{m}\sum_{i=1}^{2}\sum_{j=1}^{2}\sum_{k=1}^{2}\mu_{ijk}\left(x_{i}\ln(y_{i}+r_{ijk})+y_{j}\ln(x_{i}+r_{ijk})-z_{k}\arctan\frac{x_{i}y_{j}} {z_{k}r_{ijk}}\right)
\end{equation}
where  $x_{i}=x-L$, $y_{i}=y-L$, $z_{i}=z-L/2$,  $r_{ijk}=\sqrt{x_{i}^2+y_{j}^2+z_{k}^2}$ and $\mu_{ijk}=(-1)^{i}(-1)^{j}(-1)^{k}$.
Figure~\ref{figfig2a} represents the expected gravitational field along two lines parallel to the $z$ axis, within a range equal to $\pm20L$. The black and gray curves correspond to two vertical lines passing through points ($0, 0, 0$) and ($2L, 0, 0$), respectively. Data are normalized with the maximum vertical field $\Gamma_{max}$, occurring at point ($0, 0, -L/2$).
The FEM simulation was performed by limiting $SD$ with a sphere having a radius equal to twenty times $L$. Both the domains were meshed with an adaptive mesh generation after uniform constrain of 15 mesh elements every $L$ on the two lines where the solution is needed.
The relative difference between the simulation results and the analytical solution are shown in Figure~\ref{figfig2b}. As in the previous figure, the black curve concerns the line passing through ($0, 0, 0$) and the gray curve concerns the line passing through ($2L, 0, 0$).
The systematic error occurring near $S_{E}$ and the scattering near the body geometry are kept below 0.1~\% of the generated field. Significantly better results can be achieved by running the FEM algorithm on a platform with better memory power, which allows to increase the radius of $S_{E}$ and the number of mesh elements constrained along the lines.

\section{The cold atomic gravimeter}\label{section3}

\begin{figure}[!t]
\includegraphics[width=9in,bb=50 180 1100 500]{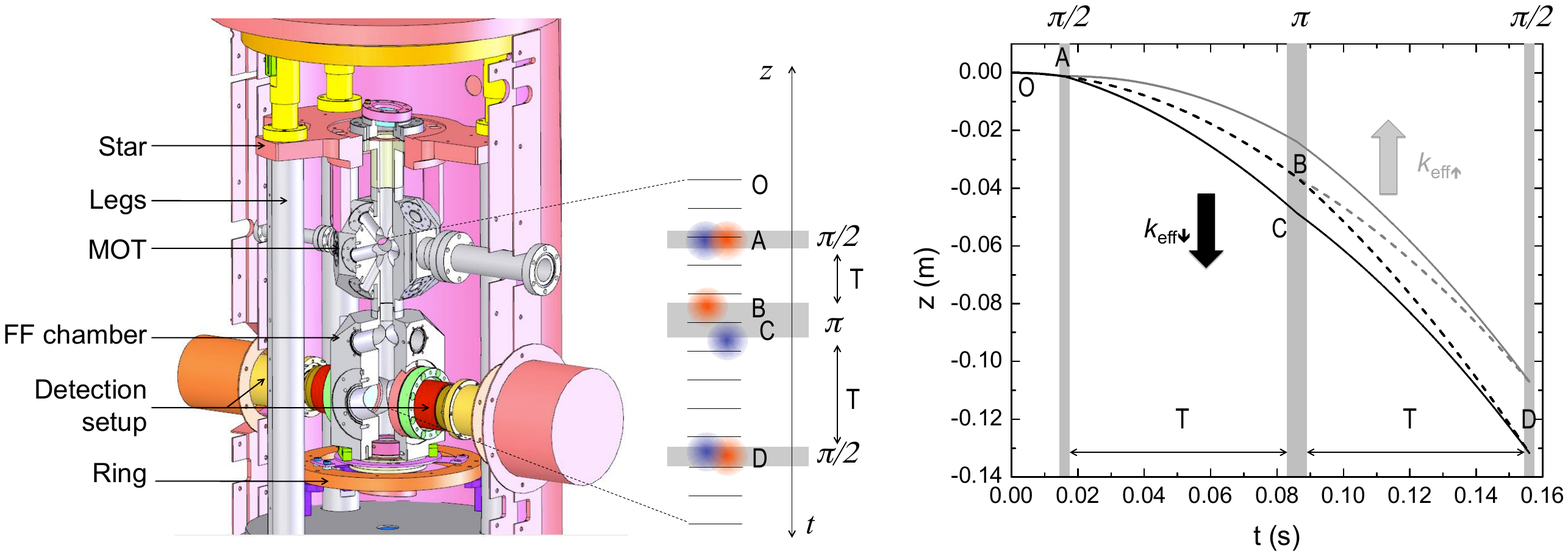}
\caption{Scheme of the interferometer. Left, vacuum chamber partly cut together with some parts of the apparatus, with the magnetic shields partially removed. Center, enlarged view of the free fall section. Right, atomic trajectories with Raman wave-vector pointing either downwards (black line) or upwards (gray lines). The Raman pulses are depicted with gray filled areas. For the sake of clarity the value of the wave-vector $k_\mathrm{eff}$ has been multiplied by 15 to distinguish the separation between path $I$ (OABD) (dash lines) and path $II$ (OACD) (continuous lines).}
\label{figgravichemins}
\end{figure}

A scheme of the CAG setup is presented in \cite{Merlet2010, Louchet-Chauvet2011}. It performs a cyclic measurement of the gravity acceleration $g$ with a cloud of $^{87}$Rb cold atoms used as a test mass~\cite{Louchet-Chauvet2011b}. The gravimeter core is shown in Figure$~$\ref{figgravichemins}.

A detailed description of the principle of the gravimeter can be found in~\cite{Louchet-Chauvet2011b}. Briefly, an atomic cloud is loaded in a 3D-MOT (Magneto Optical Trap) and is further cooled down to $2~\mu$K before being released. While the atoms fall down, three Raman pulses separated by $T$ ($\pi/2-\pi-\pi/2$) split, redirect and recombine the atomic wave packets. They are induced by two vertical counterpropagating laser beams of wave-vectors $\vec{k}_{1}$ and $\vec{k}_{2}$ which couple the hyperfine levels $\left|F=1\right\rangle$ and $\left|F=2\right\rangle$ of the $^5S_{1/2}$ ground state via a two-photon transition~\cite{Kasevich1991}. They are delivered to the atoms through a single collimator~\cite{Louchet-Chauvet2011b} and retro-reflected with a mirror placed inside the vacuum chamber. Due to conservation of angular momentum and to the Doppler shift induced by the free fall of the atoms, only two counter-propagating of the four beams will drive the Raman transitions. This feature allows to perform interferometers using effective wave-vector $\vec{k}_{\mathrm{eff}}=\vec{k}_{1}-\vec{k}_{2}$ pointing upwards or downwards. Finally, thanks to the state labelling method~\cite{Borde1989}, the interferometer phase shift $\Delta \Phi$ which is the difference of the atomic phases accumulated along the two paths $I$ and $II$ (Figure~\ref{figgravichemins}), is deduced from a fluorescence measurement of the populations of each of the two states. It is given by~\cite{Borde2001}:

\begin{equation}\label{eqdelphi}
 \Delta\Phi=\pm|\vec{k}_{\mathrm{eff}}| |\vec{g}|T^{2}
\end{equation}
where  $|\vec{k}_{\mathrm{eff}}|=|\vec{k}_{1}|+|\vec{k}_{2}|$ for counter-propagating beams. Performing the interferometer with initial atoms in state $\left|F=1\right\rangle$ leads to different paths $I$ and $II$ using $\vec{k}_{\mathrm{eff}\downarrow}$ or $\vec{k}_{\mathrm{eff}\uparrow}$ (in black and gray on Figure~\ref{figgravichemins}). The interferometer takes place in between the center of the MOT chamber ($z=0~$m) and the detection setup area ($z=-0.16~$m), through  the free-fall vacuum chamber (FF). The atoms travel in an empty vertical cylinder of $40~$mm diameter. The actual trajectory depends on the delay $t_1$ of the first pulse and on the time $T$ separating the Raman pulses. In our case, $t_1=16~$ms and $T=70~$ms, so that the interferometer occurs in the region [-1.3~mm; -120.2~mm].

\section{Mass attraction along the trajectory of the gravimeter}\label{section4}

The CAG apparatus consists of four main parts, $(i)$ the gravimeter core presented in Figure~\ref{figgravichemins}, $(ii)$ an isolation platform used to filter the vibrations due to background noise~\cite{Merlet2009}, $(iii)$ a thermal and acoustic insulation box~\cite{Legouet2008} and $(iv)$ an optical bench~\cite{Cheinet2006}.

The level of details required in the modeling of each of the four parts depends on their mass and location with respect to the atomic trajectories. Most of the attention was focused on the gravimeter core, made of several sub-components located close to the free-falling atoms. The isolation platform was modeled with an homogeneous box approximating the inner mass distribution. Due to the distance to the atoms, details of the elements inside the platform would not change the results. The insulating box was easily modeled with four lateral walls and a roof. Although the mass of the optical bench is significant, it is far away and the direction of its gravitational attraction is nearly perpendicular to the free falling motion. The numerical simulation of its effect was not necessary (maximum effect  $<10^{-10}~$m.s$^{-2}$).

Taking advantage of the linear additivity of the gravitational field, the perturbation was evaluated by splitting the whole setup on its sub-components, computing the vertical components of the field intensities along the atomÍs trajectory and adding the contributions.

All the objects and the results were modeled in a $Oxyz$ reference frame, having the origin located at the start of the atomÍs drop, with the $z$ axis vertical and upwards oriented. 

We choose to use a $S_E$ radius 100 times times larger than the maximum distance between the object surface and its center of mass, significantly larger than the factor 20 used in section~\ref{section2} for which the maximum error was as low as $0.1\%$. On the other hand, the adaptive mesh and the elements constrain along the trajectory were used depending on the magnitude of the effect. In most cases, user selection of the mesh size without element constraint were enough. from numerous tests carried out for the evaluation of the modeling uncertainty we estimate the net error well below $1~\%$ of the generated gravity field.

\begin{figure}[!t]
\includegraphics[width=7in,bb=-100 80 1500 600]{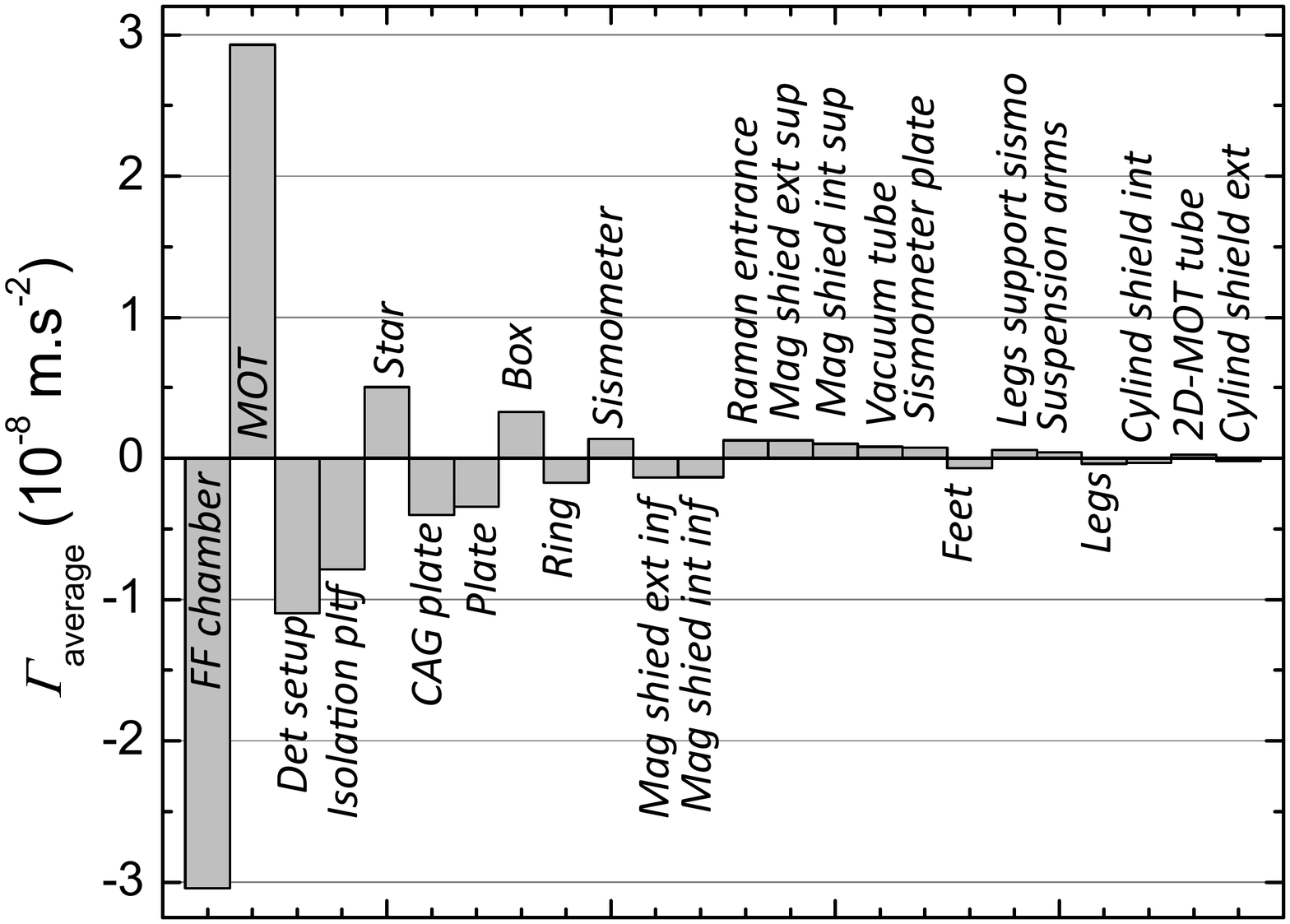}
\caption{Average gravity effect for sub-components of the CAG.}
\label{figDiagramme}
\end{figure}

\begin{figure}[!t]
\includegraphics[width=20in,bb=-100 20 1500 250]{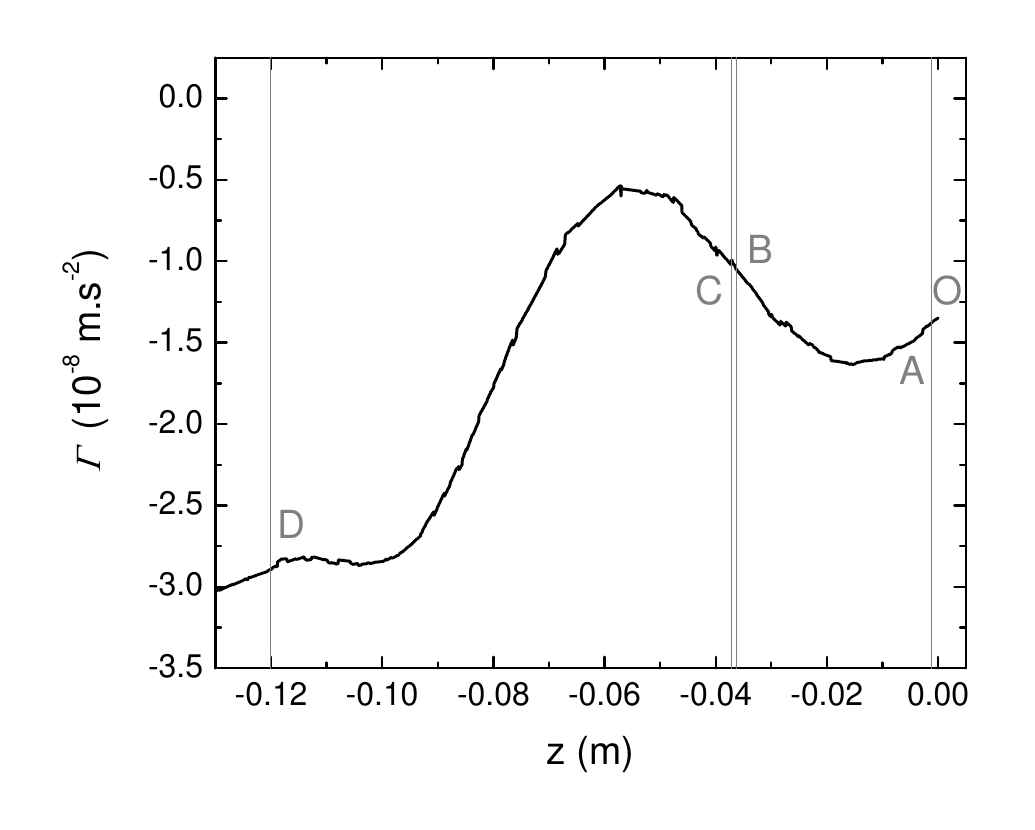}
\caption{Mass attraction of the CAG along the atomic trajectory. The cloud position at start (O) and at the three Raman pulses (A, B, C and D) are represented in gray.  }
\label{figgammaz}
\end{figure}

Figure~\ref{figDiagramme} shows the average effect for each components, from the biggest to the lowest one. The larger attraction is due to the free fall chamber  which increases the $g$ value by more than $3\times10^{-8}~$m.s$^{-2}$. The magneto optical trap (MOT) vacuum chamber compensates for most of this effect as anticipated during the design. The third component modifying $g$ with an impact larger than $10^{-8}~$m.s$^{-2}$ is the detection setup, other elements having a smaller impact. As an example, the insulating box acts upwards by about $0.3\times10^{-8}~$m.s$^{-2}$. Figure~\ref{figgammaz} displays the perturbation due to the mass attraction along the trajectory of the atoms. The parasitic acceleration is downwards oriented and varies from $1.3\times10^{-8}~$m.s$^{-2}$ to $2.9\times10^{-8}~$m.s$^{-2}$ along the atoms trajectory with a minimum of $0.5\times10^{-8}~$m.s$^{-2}$ between the second and the third Raman pulse. The effect is stronger after the last pulse, in the detection area ($<-12~$mm), and close to $3\times10^{-8}~$m.s$^{-2}$.


\section{Effect on the gravity measurement}\label{section6}

\begin{figure}[!t]
\includegraphics[width=20in,bb=-100 20 1500 250]{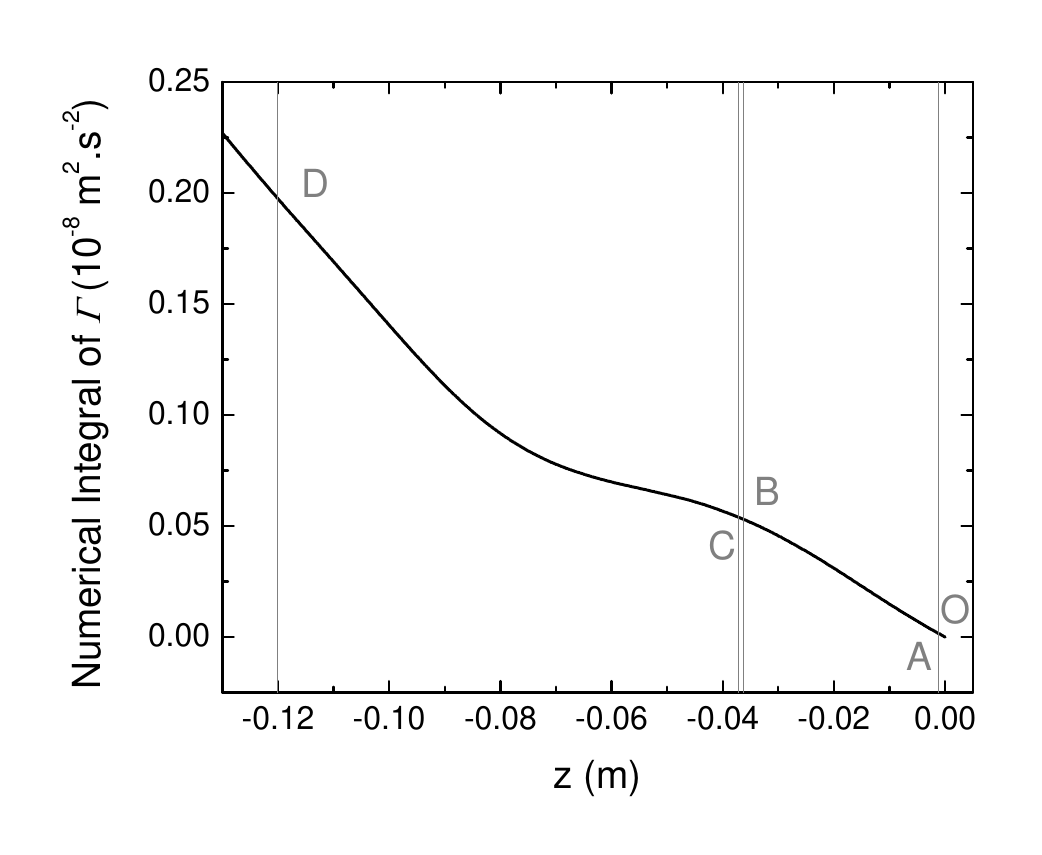}
\caption{Numerical integration of $\Gamma$ represented in Figure \ref{figgammaz}. The cloud position at start (O) and at the three Raman pulses (A, B, C and D) are represented in gray.  }
\label{figintnumgammaz}
\end{figure}

When small enough, parasitic phase shifts of the interferometer can be accurately calculated using a perturbative path integral treatment~\cite{Storey1994}. This approach already used in~\cite{Peters2001, Wolf1999} to treat the effect of a linear vertical gravity gradient, is used here to calculate the perturbation of the interferometer phase due to the mass attraction effect of the device itself ($\Delta\Phi_\mathrm{\Gamma}$). It consists on integrating the perturbed Lagrangian $\mathcal{L}_\mathrm{pert}$ along the unperturbed paths $I$ and $II$:

\begin{equation}\label{eq4}
\Delta\Phi_\mathrm{\Gamma}=\frac{1}{\hbar}\left(\int_{I}\mathcal{L}_\mathrm{pert}[z(t),\dot{z}(t)]dt-\int_{II}\mathcal{L}_\mathrm{pert}[z(t),\dot{z}(t)]dt\right)
\end{equation}
where $\hbar$ is the reduced Planck constant. In our case, the perturbed Lagrangian is linked to the perturbed potential energy which is obtained by integrating the attraction force:


\begin{equation}\label{eq7}
\mathcal{L}_\mathrm{pert}=-\delta E_{p}=m\int \Gamma (z)dz=mf(z)
\end{equation}
where $m$ is the mass of the falling atoms. Figure~\ref{figintnumgammaz}  shows the function $f(z)$ obtained integrating the mass attraction $\Gamma (z)$ plotted in Figure~\ref{figgammaz}.

The position of the falling atoms $z(t)$ is calculated for each path $I$ and $II$ taking into account the Raman pulses changing the velocity in $B$ for path $I$ and in $A$ and $C$ for path $II$, leading to the functions represented on Figure~\ref{figgravichemins}:

\begin{equation}\label{eq8}
z_{AB}(t),~z_{BD}(t),~z_{AC}(t)~\mathrm{and}~z_{CD}(t)
\end{equation}
Combining equations~\ref{eq7}~and~\ref{eq8} in equation~\ref{eq4}, $\Delta\Phi_\mathrm{\Gamma}$  can be expressed as

\begin{eqnarray*}\label{eq10}
 \Delta\Phi_\mathrm{\Gamma} & = & \frac{m}{\hbar} \left(\int_{t_1}^{t_1+T}f(z_{AB}(t))dt-\int_{t_1+T}^{t_1+2T}f(z_{BD}(t))dt \right. \\
  & & \left. -\int_{t_1}^{t_1+T}f(z_{AC}(t))dt+\int_{t_1+T}^{t_1+2T}f(z_{CD}(t))dt\right)  \\
 \end{eqnarray*}
with $t_1$ the time of the first Raman pulse. With our experimental parameters, the bias obtained with equation~\ref{eqdelphi} is $1.27 \times 10^{-8}~$m.s$^{-2}$ whatever the direction of $\vec{k}_{\mathrm{eff}}$. The difference between the two directions is negligible ($10^{-11}~$m.s$^{-2}$). Moreover, due to the finite temperature, the atoms also move radially while falling. At $2~\mu$K,  a point-like atomic cloud reaches a $1/e^2$ diameter of $4~$mm in the detection zone. We thus calculate the mass attraction $\Gamma$ for trajectories off-centered by $5~$mm in the two horizontal directions $x$ and $y$, to calculate the corresponding shifts on the $g$ measurement. We found differences with respect to the centered trajectory, as small as $4\times10^{-11}~$m.s$^{-2}$.

The global uncertainty in the calculation can be estimated by summing quadratically the uncertainties corresponding of the individual pieces. We find $5\times10^{-10}~$m.s$^{-2}$. In addition to the uncertainty in the calculation, which can be further reduced improving the mesh, we have also to account for additional uncertainties due to modeling in the mass distribution. They arise from simplifications in the geometrical description of the considered sub-components, from the influence of neglected sub-components and from imperfect knowledge of the densities. We assign a relative uncertainty of $1\%$ as a conservative estimate for these contributions which is still negligible. The total uncertainty for the self gravity error is therefore increased to $10^{-9}~$m.s$^{-2}$ which is, up to now, much lower than the targeted accuracy of the CAG.

Finally we consider the $g$ data collected with the CAG to be corrected from $(1.3 \pm 0.1)\times10^{-8}~$m.s$^{-2}$ due to the self gravity effect. This result agrees with the preliminary rough estimation of the effect of ($0\pm2)\times10^{-8}~$m.s$^{-2}$ performed for last ICAG'09~\cite{ICAG}.

\section{Conclusion}\label{conclu}

We have described a numerical method based on a FEM simulation for accurate determination of the gravitational attraction generated by any distribution of massive objects, applied to the self gravity effect on an atomic gravimeter. Using a perturbative treatment we find a relatively small systematic effect of $(1.3\pm0.1)\times10^{-8}~$m.s$^{-2}$ thanks to a symmetric design of the vacuum chamber. It shows this study was required in order to achieve the targeted accuracy.

This method allows for an accurate determination of the attraction effect of any distribution of  surrounding masses with an uncertainty arbitrary low providing densities and geometries are perfectly known. It can be applied to a large class of experiments such as watt balances~\cite{Steiner2007, Robinson2007, Bauman2009}, torsion balances~\cite{Gundlach2000, Quinn2001, Armstrong2003, Schlamminger2006} or atom interferometers for $G$ measurements~\cite{Fixler2007} and experiments testing the equivalence principle~\cite{Ertmer2009}, for which very accurate determination of the parasitic forces are mandatory in order to reach the ultimate level of performance.

\ack

The research within this EURAMET joint research project leading to these results has received funding from the European Community's Seventh Framework Programme, ERA-NET Plus, under Grant Agreement No. 217257.

\end{document}